\documentclass[preprintnumbers,aps,amsmath,amssymb]{revtex4}
\usepackage{graphicx}% Include figure files
\usepackage{dcolumn}% Align table columns on decimal point
\usepackage{bm}% bold math
\usepackage{epsfig}
\begin{document}

%\preprint{APS/123-QED}

\title{Sound absorption in glasses}% Force line breaks with \\

\author{U. Buchenau}
 \email{buchenau-juelich@t-online.de}
\affiliation{%
Forschungszentrum J\"ulich GmbH, J\"ulich Centre for Neutron Science (JCNS-1) and Institute of Biological Information Processing (IBI-8),  52425 J\"ulich, GERMANY}%
\author{G. D'Angelo, G. Carini}
\affiliation{Dipartimento MIFT, Universit$\grave{a}$ di Messina, Viale F. Stagno D'Alcontres 31, I-98166 Messina, Italy and IPCF, C. N. R., Sez. di Messina, Viale F. Stagno D'Alcontres 37, I-98158 Messina, Italy
}%
\author{X. Liu}
\affiliation{Naval Research Laboratory, Washington, DC 20375, USA
}%
\author{M. A. Ramos}
\affiliation{Laboratorio de Bajas Temperaturas, Departamento de F\'{\i}sica de la Materia Condensada, Condensed Matter Physics Center (IFIMAC)
and Instituto Nicol\'as Cabrera, Universidad Aut\'onoma de Madrid, Cantoblanco, E-28049 Madrid, Spain
}%

\date{August 5, 2022}% It is always \today, today,
             
\begin{abstract}             
The paper presents a description of the sound wave absorption in glasses, from the lowest temperatures up to the glass transition, in terms of three compatible phenomenological models. Resonant tunneling, the rise of the relaxational tunneling to the tunneling plateau and the crossover to classical relaxation are universal features of glasses and are well described by the tunneling model and its extension to include soft vibrations and low barrier relaxations, the soft potential model. Its further extension to non-universal features at higher temperatures is the very flexible Gilroy-Phillips model, which allows to determine the barrier density of the energy landscape of the specific glass from the frequency and temperature dependence of the sound wave absorption in the classical relaxation domain. To apply it properly at elevated temperatures, one needs its formulation in terms of the shear compliance. As one approaches the glass transition, universality sets in again with an exponential rise of the barrier density reflecting the frozen fast Kohlrausch $t^\beta$-tail (in time $t$, with $\beta$ close to 1/2) of the viscous flow at the glass temperature. The validity of the scheme is checked for literature data of several glasses and polymers with and without secondary relaxation peaks. The frozen Kohlrausch tail of the mechanical relaxation shows no indication of the strongly
temperature-dependent barrier density observed in dielectric data of molecular glasses with hydrogen bonds. Instead, the mechanical relaxation data indicate an energy landscape describable with a frozen temperature-independent barrier density for any glass.
\end{abstract}%\keywords{Suggested keywords}%Use showkeys class option if keyword
\maketitle

\section{Introduction}

From the point of view of elasticity, a glass is a very simple solid, elastically isotropic, described by a density $\rho$, a bulk modulus $B$ and a shear modulus $G$. Consequently, one has isotropic longitudinal and transverse sound velocities ${\rm v}_l$ and ${\rm v}_t$, respectively.

The complications begin when one looks for the absorption of these sound waves as a function of frequency and temperature. Then one realizes that there is a multitude of other excitations coexisting and interacting with the sound waves \cite{poli}.

These other excitations can be understood assuming an energy landscape \cite{cavagna}, the glass frozen not within a single structural energy minimum, but in an energy basin with many local minima. One has to reckon with low and high barriers between them, and one has to take into account the energy difference between neighboring minima of the glass. This is the conceptual basis of the tunneling model \cite{phil,varma,phillips} and its two extensions, the soft potential model \cite{klinger,bggprs,parshin,ramos,ramos1,ramos2,schober} and the Gilroy-Phillips model \cite{gilroy}.

The tunneling model \cite{phillips} is an incredibly successful empirical model at very low temperatures, with just a density of tunneling states and two coupling constants to transverse and longitudinal waves explaining a vast amount of experimental data. Its disadvantage is that it offers no clue to the fundamental question where these tunneling states come from.

The soft potential model extends the tunneling model from tunneling states to soft vibrations and low barrier relaxations in glasses, considering all three kinds of modes as members of a continuous distribution around the restoring force zero. With positive and negative restoring forces, one needs to postulate a stabilizing fourth order potential term to bring all three kinds of modes into a common picture (see Fig. 1 in reference \cite{ramos2}). There is recent numerical evidence for this postulate \cite{le1}, as well as for the increase in the number of soft vibrational modes with the fourth power of the frequency \cite{le1,le2,manning,le3,corein,mizuno,wang,le4,mizuno2,proca} which the soft potential model predicts.

The first achievement of the soft potential model is the description of the universal crossover from tunneling states to vibrations at the restoring force zero, reflected by crossovers in the temperature dependence of the specific heat and the thermal conductivity of glasses \cite{ramos} at a few Kelvin.

For the sound wave absorption, the soft potential model predicts a second universal crossover in the same temperature region, namely the one from the tunneling plateau to a classical relaxation \cite{kramers} $T^{3/4}$-rise at higher temperature. A specific aim of the paper is to quantify the decrease of the number of modes with increasing barrier height in this classical region, which leads to a peak in the sound absorption around 20 to 200 K. As will be seen, at this peak the universality of the sound absorption in glasses ends and the individuality of the specific glass begins, with relaxation peaks which are different for different glasses \cite{mazurin}.

But universality begins again as one approaches the viscous flow at the glass temperature. As known from Plazek's seminal work \cite{plazek-magill,plazek-bero,tnb,plazek-bo,plazek-pmm,roland}, the flow begins at short times $t$ with universal reversible structural relaxations, which are responsible for the fast $t^\beta$ Kohlrausch tail of the viscous flow. Plazek favors the Andrade creep value $\beta=1/3$ for the Kohlrausch exponent, but more extensive data collections \cite{bohmer,albena} show a broad scatter of values around $1/2$. 

As pointed out by Plazek \cite{charlie} and argued again in Section II. A of the present paper, any model needs to be formulated in terms of the shear compliance, in order to be able to separate the reversible Gilroy-Phillips glass relaxation processes from the irreversible viscosity processes of the liquid. Another particular aim of the present paper is to develop practical recipes within this compliance formulation for the Gilroy-Phillips-model, to determine the barrier density of reversible relaxations from measured data of a given glass and to check whether this barrier density is temperature-independent in the glass. 

A temperature-independent barrier density is the expectation of the Gilroy-Phillips model \cite{gilroy} for a constant distribution of the asymmetry of the double-well potentials around zero. Checks of this assumption with the much more powerful dielectric spectroscopy \cite{broad,gainaru,gainaru2,olsen} came to the conclusion that the Gilroy-Phillips model holds at low temperatures, but ceases to be valid at about two thirds of the glass temperature because of the strongly temperature-dependent barrier density at the Kohlrausch tail. As will be seen in the present paper, literature mechanical data for glass formers without hydrogen bonds do not support this conclusion: One does not see a temperature-dependent barrier density in the mechanical data, and the Gilroy-Phillips model holds all the way up to the glass temperature. One thus gets a continuous quantitative energy landscape description of the mechanical relaxation from the low temperature tunneling states to the onset of the flow at the glass transition.

The following Section II discusses first the general principles (II. A), then summarizes the tunneling model equations for the sound wave absorption (II. B). The soft potential equations for the sound wave absorption, their connection to the tunneling model predictions and the crossover to the low-barrier classical relaxation are detailed in II. C. Finally, the Gilroy-Phillips evaluation of sound wave absorption data by classical relaxation at higher temperatures is described in II. D. Section III applies these equations to measured data in several glasses at low (III. A) and elevated (III. B) temperatures. Section IV contains the discussion and the conclusions.

\section{Sound wave absorption in the three models}

\subsection{General considerations}

Within the glass phase, the infinite frequency moduli $G$ and $B$ have a temperature dependence which is similar to the one of the elastic moduli of the corresponding crystals, for the same reasons of anharmonicity and thermal expansion \cite{kittel}.

But what is fundamentally different from the crystalline case is that the sound absorption in glasses can almost exclusively be attributed to local structural changes. These local structural changes are structural Eshelby transformations \cite{eshelby} of an inner core of five to hundred atoms, which change the elastic misfit of the core with respect to the surrounding elastic medium. The anharmonic sound absorption known from crystals is negligible, unless one goes to very high frequencies and elevated temperatures. The influence of anharmonic processes on the sound absorption has been demonstrated for Brillouin data in vitreous silica \cite{vacher} and in alkali borate glasses \cite{cari}.  

If the barrier $V$ between the two structural states is low ($V/k_B$ 10 to 100 Kelvin), one gets a tunneling state, obeying the rules of quantum mechanics. The tunneling state can react to an external strain in two different ways, resonant or relaxational \cite{phillips}, as will be seen in detail in the next subsection.

For higher barriers $V$, the tunnel splitting becomes rapidly very small, and the two structural states are better described in a classical picture, each state in its own potential well. In this case, the transition between them occurs by classical relaxation, i.e. thermal activation over the barrier \cite{kramers} with the relaxation time of the Arrhenius equation
\begin{equation}\label{taurel}
	\tau_V=\tau_0\exp(V/k_BT).
\end{equation}
Here $\tau_0$ is usually set to a typical vibrational time of 10$^{-13}$ seconds. In the present paper, this convention will be followed for the Gilroy-Phillips model, but not for the soft potential model. For the soft potential model, we will choose the inverse of the vibration frequency in one of the wells, a better approximation for very low barriers.

Eq. (\ref{taurel}) neglects the influence of the asymmetry $\Delta$, which shortens the relaxation time by a factor $1/\cosh(\Delta/2k_BT)$ and weakens its contribution to the sound absorption by the square of the same factor. In the integration over the asymmetries, this amounts to choosing a slightly smaller $\tau_0$ and is not relevant.

After a few relaxation times, a given relaxation center in a glass has adapted to the applied external strain and does no longer contribute - unless its surroundings begin to flow. This shows that it is necessary to consider the reaction not to an external stress, but to an external strain: the theoretical treatment must not be in terms of time- or frequency-dependent moduli, but rather in terms of mechanical compliances, which describe the reaction of the sample to a given external strain. For the shear, one needs to consider the time-dependent shear compliance $J(t)$, which after the switching on of a constant shear stress $\sigma$ at the time zero leads to the time-dependent shear strain
\begin{equation}
 \epsilon(t)=J(t)\sigma,
\end{equation}
or the corresponding complex frequency-dependent shear compliance $J(\omega)$. In the shear compliance, the reversible relaxations of the glass are separated from the viscosity $\eta$ containing all irreversible processes \cite{charlie}.

For the tunneling model and the soft potential model, the distinction between moduli and compliances is not relevant, because the modulus changes only a few percent by the tunneling states and the low-barrier relaxation. But for the Gilroy-Phillips model, one needs the equations \cite{burel} for the time dependence of the shear compliance.

\subsection{Tunneling model}

The tunneling model \cite{phillips} considers tunneling states in double-well potentials with a symmetric tunnel splitting $\Delta_0=\hbar\omega_0\exp(-\lambda)$, where $\hbar\omega_0$ is the zero point energy in one of the wells and $\lambda$ is proportional to the barrier height, and an asymmetry $\Delta$, together leading to the level splitting
\begin{equation}
	E=\sqrt{\Delta_0^2+\Delta^2}.
\end{equation}

The distribution is assumed to be $P(\Delta,\lambda)=P_0$, and the couplings to the uniaxial external strain $\epsilon_l$ of the longitudinal sound waves, and the shear strain $\epsilon_t$ of the transverse sound waves, respectively, are
\begin{equation}\label{gamma}
	\gamma_l=\frac{1}{2}\frac{\partial\Delta}{\partial{\epsilon_l}}\ \ \ \gamma_t=\frac{1}{2}\frac{\partial\Delta}{\partial{\epsilon_t}}.
\end{equation}

The sound wave absorption is determined by the two dimensionless constants
\begin{equation}\label{clcttm}
C_l=\frac{P_0\gamma_l^2}{\rho v_l^2}\ \ \ C_t=\frac{P_0\gamma_t^2}{\rho v_t^2},	
\end{equation}
both of the order of 10$^{-4}$.

At low enough temperature, the lower level is markedly more populated than the upper one, and one has the resonant scattering \cite{phillips} at the frequency $\omega$
\begin{equation}\label{restun}
	Q_{res}^{-1}=\pi C_j\tanh(\hbar\omega/2k_BT),
\end{equation}
with $j=l,t$ for longitudinal and transverse waves, respectively. $Q^{-1}=\tan{\delta}$ is defined as the ratio between the imaginary and the real part of the corresponding elastic modulus at the given frequency. The resonant scattering leads to a small, but measurable sound velocity rise
\begin{equation}\label{vjt}
	\frac{v_j(T)-v_j(T_0)}{v_j(T_0)}=C_j\ln(T/T_0)
\end{equation}
where $T_0$ is an arbitrary low reference temperature.

One has to go to large frequencies (order of GHz) and low temperatures (below 100 mK) so that $\hbar\omega>k_BT$ in order to see the resonant scattering. But in this range one is able to identify the two-level character of the excitations \cite{hunkli}: Making the microwave intensity $I$ high enough, the sound absorption disappears, because the two levels become equally populated. Instead of eq. (\ref{restun}), one then finds
\begin{equation}\label{restuns}
	Q_{res}^{-1}=\pi C_j\frac{\tanh(\hbar\omega/2k_BT)}{\sqrt{1+I/I_{c1}}},
\end{equation}
where $I_{c1}$ is a critical intensity of the order of 10$^{-7}$ W/cm$^2$.

One has not only the resonant response for a given tunneling state, but also the relaxational one, given by the relaxation time $\tau_t$ of the tunneling state to adapt to the distortion of the sound wave. Unlike the resonant scattering, the relaxational scattering is not intensity-dependent, because it stems from a much broader distribution of tunneling states. $\tau_t$ is determined by the interaction of the tunneling state with all sound waves
\begin{align}
	\tau_t^{-1}=\frac{E\Delta_0^2\coth(E/2k_BT)}{2\pi\rho\hbar^4}\left(\frac{\gamma_l^2}{{\rm v}_l^5}+\frac{2\gamma_t^2}{{\rm v}_t^5}\right).
\end{align}

At very low temperatures, the relaxational sound absorption rises proportional to $T^3$ and saturates at the tunneling plateau value 
\begin{equation}\label{reltun}
	Q_{rel}^{-1}=\frac{\pi C_j}{2}.
\end{equation}

Taking only the relaxational sound absorption into account, one derives the opposite behavior of the sound velocity to the one of eq. (\ref{vjt}) 
\begin{equation}\label{vjtr}
	\frac{v_j(T)-v_j(T_0)}{v_j(T_0)}=-\frac{3}{2}C_j\ln(T/T_0).
\end{equation}
Adding the two sound velocity changes of equs. (\ref{vjt}) and (\ref{vjtr}), one obtains first a logarithmic rise with $C_j$ and then a decrease with $-C_j/2$ (the sum of both effects) in the plateau region. 

These results are nearly identical in the soft potential model and will be discussed in more detail in the next subsection.

\subsection{Soft potential model}

\subsubsection{Analytical approximations}

The soft potential model \cite{klinger,bggprs,parshin,ramos,ramos1,ramos2,schober} is more detailed than the tunneling model. It assumes the knowledge of the potential in the normal coordinate $A$ of the soft modes, defined by the kinetic energy $E_{kin}=\dot{A}^2/2$, allowing a numerical calculation of the splitting $\Delta_0$. It is not identical with the tunneling model, because its assumptions on density and coupling are different. Its prediction for the resonant sound wave scattering from the tunneling states is the same as in the tunneling model, but its prediction for the relaxational sound wave scattering is only the same at frequencies below 1 MHz, as we will demonstrate in the second part of the subsection.

The soft potential model is based on the concept of a coexistence of the sound waves with localized low-frequency modes. In the model, the potential of these modes has a uniform stabilizing fourth-order term. In addition, each mode has its individual first-order asymmetry and second-order restoring force constant terms, which can be either positive or negative. Thus one can have single-well or double-well potentials with different degrees of asymmetry.

The model postulates a constant density
\begin{equation}\label{psimp}
	P(D_1,D_2)=P_s
\end{equation}
of soft modes in the $D_1-D_2$-plane of the potential equation
\begin{equation}
	V(x)=W(D_1x+D_2x^2+x^4)
\end{equation}
around the purely quartic potential with $D_1=D_2=0$ and the zero point energy $W$, with the dimensionless coordinate $x$ defined in such a way that the zero point energy becomes the prefactor of the fourth order term.

$W$ is an energy of a few tenths of one meV. This zero-point energy is defined by the quantum mechanical equilibrium of kinetic confinement and potential energy for the wave function of the lowest energy level in the purely quartic potential. In terms of the normal coordinate $A$, the potential prefactor is $D_4$, and the mean square displacement $A_0^2$ of the ground level is given by
\begin{equation}\label{zp}
	D_4A_0^4=\frac{\hbar^2}{2A_0^2}\equiv W
\end{equation}
leading to
\begin{equation}\label{wa0}
	W=\frac{\hbar^{4/3}D_4^{1/3}}{2^{2/3}}\ \ \ A_0=\frac{\hbar^{1/3}}{2^{1/6}D_4^{1/6}}.
\end{equation}
The dimensionless coordinate is $x=A/A_0$.

For a given negative $D_2$ and $D_1=0$, one gets a symmetric double-well potential with the barrier height
\begin{equation}\label{V}
	V=W\frac{D_2^2}{4}.
\end{equation}
It is possible to calculate the tunnel splitting $\Delta_0$ for this potential numerically as a fraction of $W$.

On the positive force constant side, the soft potential model predicts a vibrational density of soft modes increasing with the fourth power of the frequency \cite{klinger,bggprs,parshin,ramos,ramos1,ramos2,schober}, the prediction corroborated by recent numerical evidence \cite{le1,le2,mizuno,wang}. But this vibrational density of states does not lead to any sound wave absorption below the frequency $W/h$ of about 50 GHz, which separates tunneling states and soft vibrations.

Therefore one can restrict the treatment to asymmetric double-well potentials with a barrier height $V$ and the asymmetry $\Delta$ like the one in Fig. 1, which one finds at negative $D_2$-values for $\left| D_1\right|$ smaller than the limiting value
\begin{equation}\label{d1lim}
	D_{1l}=\frac{4\left|D_2\right|^{3/2}}{3\sqrt{6}}.
\end{equation}

%%%%%%%%%%%%%%%%%%%%% begin figure %%%%%%%%%%%%%%%%%%%%%%%%%%%%%%%%%%%%%
\begin{figure}[t]
\hspace{-0cm} \vspace{0cm} \epsfig{file=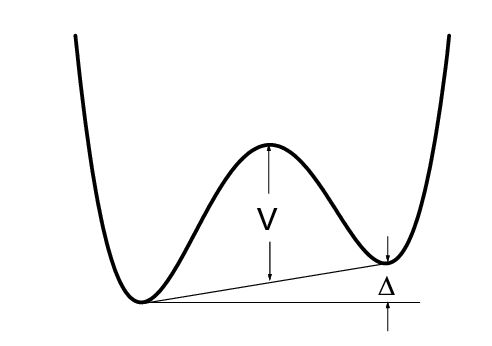,width=6 cm,angle=0} \vspace{0cm} \caption{Definition of the barrier height $V$ and the asymmetry $\Delta$ of an asymmetric double well potential.}
\end{figure}
%%%%%%%%%%%%%%%%%%%%% end figure %%%%%%%%%%%%%%%%%%%%%%%%%%%%%%%%%%%%%%%

Soft potential fits \cite{ramos} of the low temperature anomalies of glasses indicate a crossover energy $W$ corresponding to a thermal energy of a few Kelvin. This implies that the tunneling states responsible for the low temperature anomalies below 1 K lie between $D_2=-13$ and $D_2=-3$. In this range, the quasiclassical Wentzel-Kramers-Brillouin-(WKB)-approximation for the tunnel splitting $\Delta_0$ and the asymmetry $\Delta$ of the tunneling states is not good enough to reach the 10 percent level which one needs for a comparison to experiment. For this reason, numerical calculations were done \cite{ramos1}, leading to the approximations
\begin{equation}\label{delta0}
	\Delta_0=W(-D_2)^{3/2}\exp(1-\sqrt{2}(-D_2)^{3/2}/3),
\end{equation}
reaching the $\mu$K  range at $D_2=-13$, and
\begin{equation}\label{deltat}
	\Delta=WD_1\sqrt{2(-D_2-1)}.
\end{equation}

The interaction between tunneling states and sound waves is determined by the bilinear coupling energies of the soft potential model \cite{ramos,ramos1}
\begin{equation}
	\delta V_l=\Lambda_lx\epsilon_l\ \ \ \delta V_t=\Lambda_tx\epsilon_t,
\end{equation}
where $\epsilon_l$ is the uniaxial strain of a longitudinal sound wave and $\epsilon_t$ is the shear strain of a transverse sound wave. $\Lambda_l$ and $\Lambda_t$ are the corresponding coupling constants. The relation to the coupling constants $\gamma_l$ and $\gamma_t$ of the tunneling model \cite{phillips} is given by
\begin{equation}\label{gammat}
 \gamma_t=\frac{1}{2}\frac{\partial\Delta}{\partial\epsilon_t}=\Lambda_t\sqrt{(-D_2-1)/2}
\end{equation}
and the corresponding equation for $\gamma_l$.

With these equations, one can define soft potential values $C_l$ and $C_t$ corresponding to those of the tunneling model in eq. (\ref{clcttm})
\begin{equation}\label{c}
	C_l=\frac{P_s\Lambda_l^2}{W\rho v_l^2}\ \ \ C_t=\frac{P_s\Lambda_t^2}{W\rho v_t^2}
\end{equation}
and finds that one gets again the equations (\ref{restun},\ref{vjt},\ref{reltun}) for the sound absorption of the tunneling model, though with a slight modification: The equations have to be multiplied with
\begin{equation}\label{fd2}
	f(D_2)=\frac{\sqrt{-D_2-1}}{\sqrt{-D_2}+3/\sqrt{2}D_2},
\end{equation}
where $D_2$ defines the average tunneling state which is seen by the sound wave with the given frequency at the given temperature. But since $f(D_2)$ only varies from 1.01 to 1.07 between $D_2$ from $-13$ to $-6$, the soft potential model has practically the same predictions for the sound wave absorption as the tunneling model in a wide range of frequencies and temperatures \cite{ramos1}.

At higher temperatures, one can no longer neglect the classical Arrhenius relaxation by thermally activated jumps. Integrating over all relevant $D_1$- and $D_2$-combinations, the classical relaxation leads to the sound absorption \cite{ramos} 
\begin{equation}\label{qclass}	
	Q^{-1}_{rel,class}=\frac{\pi C_j}{\ln(1/\omega\tau_0)^{1/4}}\left(\frac{k_BT}{W}\right)^{3/4},
\end{equation}
which rises with $T^{3/4}$.

Note that the $T^{3/4}$-rise comprises two increases: a barrier density increase with $V^{1/4}$, recently corroborated numerically \cite{le4}, which follows from equs. (\ref{V}) and (\ref{d1lim}), and the coupling constant increase with $V^{1/4}$, which follows from eq. (\ref{gammat}) for large $D_2$.

In the crossover region, a given tunneling state has two competing possibilities to adapt to the external shear distortion of a sound wave, either by the tunneling lifetime or by a thermally activated jump over the barrier. At low temperature, the tunneling lifetime decay dominates, at high temperature the thermal activation. Note that both processes become exponentially slower with increasing barrier height, but only the thermal activation has the exponent $-V/k_BT$ in the decay rate.

The simplest description \cite{ramos} of this situation is to define a crossover temperature $T_c$, where both contributions are equal. Below $T_c$, one sets the sound absorption equal to the tunneling contribution of eq. (\ref{reltun}), above to the classical relaxation value of eq. (\ref{qclass}). The crossover temperature $T_c$ is
\begin{equation}\label{kbtc}
	k_BT_c=\frac{\ln(1/\omega\tau_0)^{1/3}}{2^{4/3}}W.
\end{equation}
For the usual assumption $\tau_0=10^{-13}$ s, this implies $k_BT_c=1.206\ W$ at a frequency of 1 Hz and $k_BT_c=0.96\ W$ at 1 MHz, so the crossover temperature lies close to $W/k_B$. This crossover approximation has been shown to be valid in several cases \cite{ramos}.

But there are situations where the approximation fails. In these cases, it is necessary to integrate numerically over all states, with appropriate assumptions on the two possible ways in which a given state can adapt to the distortion of the sound wave. This will be done in the next subsection.  

\subsubsection{Numerical crossover calculation} 

The coupling to the sound waves determines the lifetime $\tau_t$ of the tunneling states. Rewriting the expression for the relaxation time of the tunneling model \cite{phillips} in terms of the parameters of the soft potential model \cite{ramos1}, one gets
\begin{equation}\label{taut}
	\tau_t^{-1}=A_t\frac{-D_2-1}{2}\Delta_0^2E\coth(E/2k_BT)
\end{equation}
with
\begin{equation}\label{at}
	A_t=\frac{1}{2\pi\rho\hbar^4}\left(\frac{\Lambda_l^2}{{\rm v}_l^5}+\frac{2\Lambda_t^2}{{\rm v}_t^5}\right).
\end{equation}

The tunneling decay rate of eq. (\ref{taut}) competes with the classical relaxation by the thermal activation to states lying above the barrier \cite{kramers}, leading to the Arrhenius decay rate of eq. (\ref{taurel}). In this equation, the attempt rate $\tau_0^{-1}$ is given by the vibration frequency $\nu$ in the wells for a symmetric double-well potential
\begin{equation}\label{tau0}
 h\nu=2W\sqrt{-2D_2}.	
\end{equation}

But eq. (\ref{taurel}) begins to fail at low tunneling barriers, where it predicts a faster decay than the real one.

To see this, remember that one considers a transition from one of the two lowest levels of the double well potential to an excited level above the barrier, from which it returns to the other lowest level. The situation is illustrated for a soft potential with $D_1=0.1$ and $D_2=-6$ in Fig. 2 (a). In this case, the assumption that the energy needed for the transition is the barrier height is an overestimate: the barrier height is 9 W, while the excitation energy to the level above the barrier is only 6.4 W.

But consider the case of the purely quartic potential in Fig. 2 (b), where the barrier has gone down to zero. In this case, the direct transition from the ground state to the first excited level is the one with the tunneling relaxation time $\tau_t$. The thermally activated relaxation time $\tau_{rel}$ corresponds to the excitation of the second excited level, which is 6.45 W higher than the ground state, and then to the return to the first excited level.

%%%%%%%%%%%%%%%%%%%%% begin figure %%%%%%%%%%%%%%%%%%%%%%%%%%%%%%%%%%%%%
\begin{figure}[t]
\hspace{-0cm} \vspace{0cm} \epsfig{file=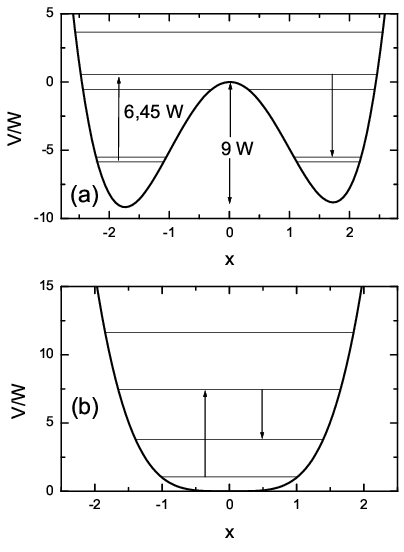,width=6 cm,angle=0} \vspace{0cm} \caption{Thermally activated transition between the lowest two energy levels in (a) the soft potential with $D_1=0.1$ and $D_2=-6$ (b) the purely quartic potential with $D_1=D_2=0$.}
\end{figure}
%%%%%%%%%%%%%%%%%%%%% end figure %%%%%%%%%%%%%%%%%%%%%%%%%%%%%%%%%%%%%%%

A crude approximation to take this effect into account is to replace the real barrier $V$ of the double well potential in eq. (\ref{taurel}) by 6.5 $W$, whenever $V$ is smaller. We will see that the crossover to this approximation is the point where the soft potential predictions begin to differ from the tunneling model ones.

In order to enter this domain, it is necessary to consider tunneling states in the range between $D_2=-6$ and $D_2=-1$, where eq. (\ref{delta0}) fails and has to be replaced by
\begin{equation}
	\log(\Delta_0/W)=0.438+0.16D_2+0.015D_2^2+0.006D_2^3.
\end{equation}

Since the two decay processes, the direct transition to the other tunneling level and the indirect transition via the excitation to a higher level, are independent, the total decay rate is the sum of the two decay rates
\begin{equation}
	\tau^{-1}=\tau_t^{-1}+\tau_V^{-1}.
\end{equation}

A two-level state with the energy splitting $E$ and the coupling $\gamma_t$ to the shear strain $\epsilon_t$ has the free energy
\begin{equation}
	F=-k_BT\ln\left[2\cosh\left(\frac{E+2\gamma_t\epsilon_t}{2k_BT}\right)\right].
\end{equation}

Its contribution to the relaxational reduction of the shear modulus $G$ at times much longer than its relaxation time is given by the second derivative
\begin{equation}
	\frac{\partial^2F}{\partial\epsilon_t^2}=\frac{\gamma_t^2}{k_BT\cosh^2(E/2k_BT)}
\end{equation}

Knowing the splitting $E$ of the two states and their coupling to an external shear distortion, it is straightforward to calculate the contribution $\delta G(D_1,D_2)$ to the weakening of the shear modulus $G$ from the modes in the small square $dD_1dD_2$ between $D_1$ and $D_1+dD_1$ as well as $D_2$ and $D_2+dD_2$
\begin{equation}
	\delta G(D_1,D_2)=A(D_1,D_2)dD_1dD_2,
\end{equation}
with
\begin{equation}
	A(D_1,D_2)=\frac{\Lambda_t^2}{2k_BT}\frac{(-D_2-1)P_s(D_1,D_2)}{\cosh^2(E/2k_BT)}\frac{\Delta^2}{E^2}.
\end{equation}

Integrating over $D_1$ and $D_2$, one finds \cite{zacc} the complex shear modulus $G(\omega)$ at the frequency $\omega$
\begin{equation}\label{gp}
	G'(\omega)=G-\int_{-D_{1l}}^{D_{1l}} dD_1\int_{-\infty}^{-1}dD_2\frac{\omega^2\tau^2A(D_1,D_2)}{1+\omega^2\tau^2}
\end{equation}
and
\begin{equation}\label{gpp}
	G''(\omega)=\int_{-D_{1l}}^{D_{1l}} dD_1\int_{-\infty}^{-1}dD_2\frac{\omega\tau A(D_1,D_2)}{1+\omega^2\tau^2}.
\end{equation}
Here the integral over $D_2$ is only extended to $D_2=-1$, where the approximation for the coupling coefficient, eq. (\ref{gammat}), extrapolates to zero.

Equs. (\ref{gp}) and (\ref{gpp}) calculate exclusively the relaxational scattering, so one has to add the resonant scattering from the tunneling states of eq. (\ref{restun}) \cite{phillips}. However, this is only relevant at high frequencies and very low temperatures.

In the same way, exchanging $\Lambda_t$ with $\Lambda_l$, one can calculate the complex longitudinal modulus $M(\omega)$ of the longitudinal sound waves.

%%%%%%%%%%%%%%%%%%%%% begin figure %%%%%%%%%%%%%%%%%%%%%%%%%%%%%%%%%%%%%
\begin{figure}[t]
\hspace{-0cm} \vspace{0cm} \epsfig{file=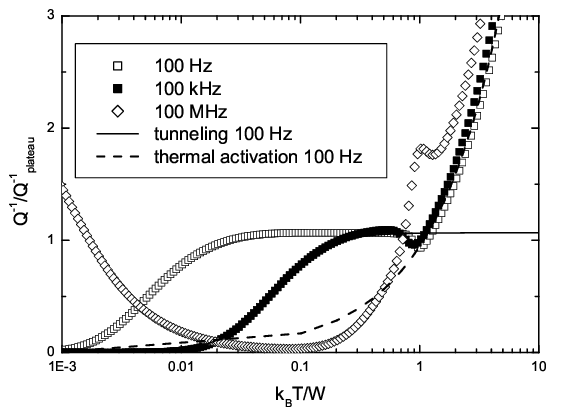,width=6 cm,angle=0} \vspace{0cm} \caption{Theoretical sound absorption for 100 Hz, 100 kHz and 100 MHz with the parameters for vitreous silica from Table I. Note the rise of the 100 MHz results over the plateau value.}
\end{figure}
%%%%%%%%%%%%%%%%%%%%% end figure %%%%%%%%%%%%%%%%%%%%%%%%%%%%%%%%%%%%%%%

Fig. 3 shows calculated sound absorption curves in terms of $Q^{-1}=G''/G'$ for 100 Hz, 100 kHz and 100 MHz, using the parameters for vitreous silica in Table I. Note that only the high frequency shows the strong resonant tunneling absorption of eq. (\ref{restun}) at very low temperature. For the two lower frequencies, the resonant contribution is negligible, and one finds at low temperature the $T^3$-rise of the tunneling model
\begin{equation}
	Q^{-1}=\frac{\pi^4C_t(-\overline{D_2}-1)}{24\omega}A_t(k_BT)^3.
\end{equation}
In this case, $\overline{D_2}$ is determined by the condition $\omega\tau_t=1$ with eq. (\ref{taut}) at the low temperature, for 100 Hz in silica $\overline{D_2}=-8$.

In the 100 Hz and 100 kHz calculation, one finds the saturation to the plateau, slightly higher than the tunneling plateau of eq. (\ref{reltun}) because of the factor $f(D_2)$ of eq. (\ref{fd2}). Then there is a slight dip and then the crossover to classical relaxation. In these two cases, the approximation \cite{ramos} to take only the tunneling below the crossover and only the relaxation above is supported by the numerical results.

This is no longer true for the 100 MHz calculation, where the rise to saturation coincides with the crossover to classical relaxation. Here, one finds a peak at the crossover, which is higher than the plateau. The peak height increases with increasing frequency (see Figs. 4 and 5 in Section III. A). Looking into the calculation, one finds that the tunneling states relevant for the relaxational response begin to approach a $\Delta_0$-value close to $W$, so the barrier for the thermally activated relaxation via higher energy levels begins to be larger than the value $WD_2^2/4$ calculated from eq. (\ref{V}). This causes the rise over the tunneling plateau. 

Note that the relevant $\Delta_0$-value for the resonant scattering at 100 MHz is still a factor of fifty lower than $W$. Therefore the resonant scattering remains the one of the tunneling model up into the GHz range; it is only the relaxational response which suggests a higher plateau value.

Eq. (\ref{gp}) allows to calculate the relaxational change of the transverse sound velocity, to which one can add the resonant sound velocity change of eq. (\ref{vjt}), which is the same in both models. The corresponding procedure for the longitudinal modulus supplies the change of the longitudinal sound velocity (see Fig. 5 (b) in Section III. A).

Note also that the present treatment is rather simplistic, neglecting possible complications like the multiphonon effects on the tunneling states \cite{grascho} or their possible elastic interaction \cite{burin}. 

To summarize, the soft potential model predicts the same resonant sound wave scattering and the same low frequency relaxational plateau from tunneling states as the tunneling model, but the sound absorption rises over the plateau at higher frequencies, when the responsible tunneling states approach the crossover to vibrations. At frequencies below 1 MHz, the simple sketchy derivation \cite{ramos} of the crossover temperature from tunneling to relaxation of eq. (\ref{kbtc}) holds. 

The comparison to experimental data in Section III will show examples for the increase of the tunneling plateau at higher frequencies. Also, it will be seen that the relaxational rise at higher temperatures does not go on forever. In many glasses, one can describe this by replacing the constant distribution in $D_1,D_2$ by a gaussian around $D_2=0$
\begin{equation}\label{gau}
	P(D_1,D_2)=P_s\exp(-bD_2^2W/4k_BT_g),
\end{equation}
where $b$ is adapted to the peak in the sound absorption which one finds around 100 K ($b=1$ means that at the barrier of height $k_BT_g$ the probability to find it is reduced by $1/e$).

\subsection{Gilroy-Phillips model}

As shown in the preceding subsection, the sound absorption mechanism changes from tunneling to classical relaxation at a crossover temperature $T_c$, estimated \cite{ramos} to lie at 1.2 $W/k_B$.

High enough above $T_c$, one can describe all relaxations in terms of the classical Kramers-Arrhenius relaxation mechanism in asymmetric double-well potentials like the one in Fig. 1 \cite{kramers}. In a close parallel to the tunneling model and the soft potential model, the Gilroy-Phillips model \cite{gilroy} assumes a constant distribution of asymmetries $\Delta$ around zero.

Integrating over the asymmetry, one finds the classical shear relaxation of the glass in terms of a temperature-independent barrier density function $l(V)$, given by \cite{burel}
\begin{equation}\label{lv}
	l(V)=\frac{4\gamma_t^2n(V,0)}{G},
\end{equation}
where $n(V,\Delta)$ is the number density of relaxing entities with barrier height V and the asymmetry $\Delta$ (see Fig. 1), and the coupling constant $\gamma_t$ is defined by the tunneling model eq. (\ref{gamma}). The factor 4 in eq. (\ref{lv}) ensures that one gets a shear compliance increase $GdJ=l(V)dV$ for relaxations between $V$ and $V+dV$. This increase is temperature-independent as long as the energy landscape of the glass remains temperature-independent. A single symmetric double-well potential causes a $dJ\propto 1/T$, but the integration over the asymmetries cancels this factor. As a consequence, one can describe the classical relaxation with a temperature-independent barrier density. That is the great advantage of the Gilroy-Phillips model.

The time-dependent shear compliance $J(t)$ is given by \cite{burel}
\begin{equation}\label{jt}
	GJ(t)=1+\int_0^\infty l(V)(1-\exp(-t/\tau_V))dV+t/\tau_M,
\end{equation}
and the frequency-dependent complex shear compliance $J(\omega)=1/G(\omega)$ is
\begin{equation}\label{jom}
	GJ(\omega)=1+\int_0^\infty \frac{l(V)dV}{1+i\omega\tau_V}-\frac{i}{\omega\tau_M}.
\end{equation}
Here $\tau_V$ is the Arrhenius relaxation time of eq. (\ref{taurel}) and $\tau_M$, the Maxwell time $\eta/G$ ($\eta$ viscosity), is infinite in a glass, so the last term of both equations need not be taken into account. Note that $G$ is the infinite frequency shear modulus in both equations.

To determine the barrier density $l(V)$ of a given glass from sound absorption measurements, one makes the approximation \cite{burel} that $l(V)$ varies so slowly with the barrier height that it can be taken to be constant over more than a decade of relaxation times at the given frequency $\omega$ and temperature $T$. With $V(\omega,T)=k_BT\ln(1/\omega\tau_0)$, this approximation yields
\begin{equation}\label{gjp}
	GJ(1/\omega)=GJ'(\omega)=\int_0^{k_BT\ln(1/\omega\tau_0)}l(V)dV
\end{equation}
and
\begin{equation}
	GJ''(\omega)=-\frac{\pi}{2}l(k_BT\ln(1/\omega\tau_0)))k_BT.
\end{equation}

If a measurement supplies $Q^{-1}=\tan{\delta}=-J''/J'$, one has
\begin{equation}\label{lvq}
	l(k_BT\ln(1/\omega\tau_0))=\frac{2GJ'(\omega)Q^{-1}}{\pi k_BT},
\end{equation}
so one needs not only $Q^{-1}$, but also $GJ'(\omega)$ to determine $l(V)$.

At low temperature, one can reckon with $GJ'(\omega)=1$ for all frequencies. Depending on the glass, the value rises to a value around 2 as one approaches the glass temperature.

For a measurement at constant frequency which starts at low temperature, one starts with $GJ'(\omega)=1$ at the first measured point. The measured point provides the $l(V)dV$ to calculate $GJ'(\omega)$ for the next measured point. Thus one can integrate over the temperatures to get $GJ'(\omega)$ for each measured point. One does not even need $G$ then, and can neglect the temperature dependence of $G$ if $l(V)$ is indeed temperature-independent. If it is, measurements at different frequencies supply the same result. We will come back to this point in the discussion in Section IV.

To compare with $l(V)$ in the equilibrium undercooled liquid at the glass temperature, one can use Plazek's measurements of the recoverable compliance $J_r(t)=J(t)-t/\eta$ {\it vs} $\log{t}$ (the logarithm to the basis 10), to obtain $l(V)$ from
\begin{equation}\label{jtlv}
	l(2.303k_BT(\log{t}+13))=G\frac{\partial J_r(t)}{\partial\log{t}}\frac{1}{2.303k_BT}.
\end{equation}
In this case, one needs $G$ at the given temperature from an ultrasonic or Brillouin measurement.

To determine $l(V)$ from a measurement of $G(\omega)$ at the glass temperature, one inverts $J(\omega)=1/G(\omega)$, subtracts the viscous part from $J''(\omega)$ and uses eq. (\ref{lvq}). $G$ and $\eta$ are needed.
 
The absorption for a shear distortion is described by
\begin{equation}\label{def}
Q^{-1}=\tan{\delta}=-\frac{J''}{J'}=\frac{{\rm v}_tl_{rel,class}^{-1}}{\omega},
\end{equation}
where ${\rm v}_t$ is the transverse sound velocity, $l_{rel,class}^{-1}$ is the inverse mean free path of the transverse sound waves under the influence of classical relaxation and $\omega$ is the angular frequency of the sound wave.
Oscillator data supply $Q^{-1}$, the sound absorption data are given as $\alpha_{db}$ in db/cm, from which one can calculate $Q^{-1}$ via \cite{poli}
\begin{equation}
	Q^{-1}=0.23\frac{{\rm v}}{\omega}\alpha_{db},
\end{equation}
where ${\rm v}$ is the respective sound velocity.

For longitudinal sound absorption data one should replace $J''/J'$ by the corresponding expressions for a longitudinal strain deformation, defining a longitudinal $l(V)$. But the comparison to experiment will show that these two $l(V)$ are usually nearly equal.

For the simplest form of the soft potential model with a constant density $P_s$ of soft modes in the $D_1-D_2$-plane, one finds \cite{burel}
\begin{equation}
	l_j(V)=\frac{2C_j}{V^{1/4}W^{3/4}},
\end{equation}
where $j=l,t$, for longitudinal and transverse sound waves, respectively.

This soft potential barrier density proportional to $V^{-1/4}$ seems to contradict the earlier statement in Section II. C, according to which the barrier density increases with $V^{1/4}$, in agreement with the numerical finding \cite{le4}. But the difference lies in the definition of $l(V)$ via $n(V,0)$, which still has to be integrated over the asymmetry $\Delta$ to obtain the full barrier density.

\section{Comparison to experiment}

\subsection{The crossover from tunneling to classical relaxation}

The first two examples, vitreous silica and germania, corroborate the soft potential prediction of Fig. 3.  In these two and the two following glasses, the soft potential parameters were taken from the fit \cite{ramos} of heat capacity and thermal conductivity data, demonstrating again the impressive consistency of the soft potential model \cite{ramos}.

%%%%%%%%%%%%%%%%%%%%% begin figure %%%%%%%%%%%%%%%%%%%%%%%%%%%%%%%%%%%%%
\begin{figure}[t]
\hspace{-0cm} \vspace{0cm} \epsfig{file=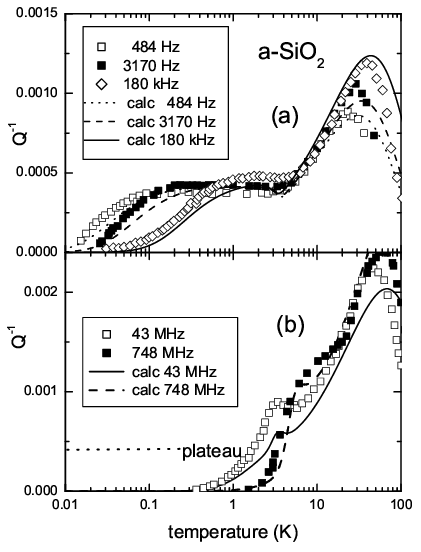,width=7 cm,angle=0} \vspace{0cm} \caption{Comparison of measured low temperature sound absorption data with the soft potential calculation in vitreous silica (parameters see Table I) (a) flexural data at 484 and 3170 Hz \cite{ray} and at 180 kHz \cite{keil} (b) longitudinal ultrasonic data at 43 MHz \cite{bartell} and 748 MHz \cite{jones}, where both data and soft potential calculation rise above the low frequency plateau, though the calculated rise at 43 MHz is weaker than the measured one.}
\end{figure}
%%%%%%%%%%%%%%%%%%%%% end figure %%%%%%%%%%%%%%%%%%%%%%%%%%%%%%%%%%%%%%%

Fig. 4 compares the soft potential calculations to vitreous silica data \cite{ray,keil,bartell,jones}, in Fig. 4 (a) to low frequency data where soft potential and tunneling predictions agree and in Fig. 4 (b) to high frequency data where they do not agree.

Taking the values $P_s$, $W$ and the average $C$ in Table I, one can calculate $\Lambda_l$ and $\Lambda_t$ from the equations (\ref{c}) under the assumption that the two $C$-values are equal. One then has everything needed to calculate the sound absorption from the equations in Section II. C. The calculation reproduces the low frequency flexural data \cite{ray,keil} in Fig. 4 (a), which follow the tunneling model predictions. The higher frequency longitudinal ultrasonic data \cite{bartell,jones} in Fig. 4 (b) show the increase of the plateau height, first noted by Topp and Cahill \cite{topp}. In the 43 MHz data \cite{bartell}, this rise is markedly stronger than the soft potential prediction, but for the higher frequency one finds good agreement, so one can conclude that the rise of the tunneling plateau is indeed consistent with the soft potential model, though the model predicts it at a somewhat higher frequency. 

%%%%%%%%%%%%%%%%%%%%% begin figure %%%%%%%%%%%%%%%%%%%%%%%%%%%%%%%%%%%%%
\begin{figure}[t]
\hspace{-0cm} \vspace{0cm} \epsfig{file=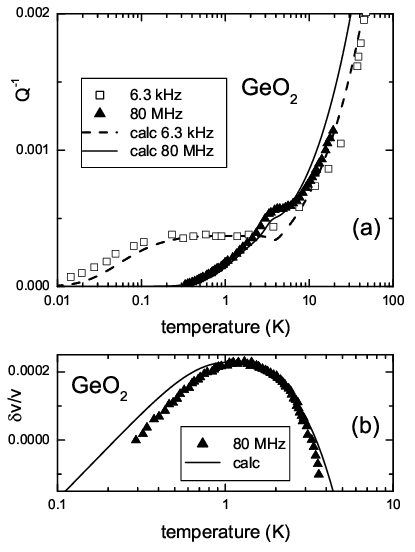,width=7 cm,angle=0} \vspace{0cm} \caption{(a) Comparison of measured low temperature flexural absorption data in vitreous germania at 6.3 kHz \cite{rau} and for longitudinal ultrasonic data \cite{laermans} at 80 MHz with the soft potential calculation (parameters see Table I). The MHz data show again a higher plateau, in agreement with the soft potential prediction (b) Comparison of the sound velocity changes in the same 80 MHz data \cite{laermans} with the soft potential calculation.}
\end{figure}
%%%%%%%%%%%%%%%%%%%%% end figure %%%%%%%%%%%%%%%%%%%%%%%%%%%%%%%%%%%%%%%

%%%%%%%%%%%%%%%%%%%%% begin figure %%%%%%%%%%%%%%%%%%%%%%%%%%%%%%%%%%%%%
\begin{figure}[b]
\hspace{-0cm} \vspace{0cm} \epsfig{file=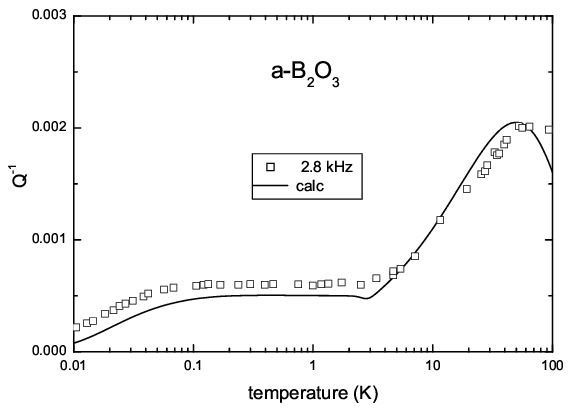,width=7 cm,angle=0} \vspace{0cm} \caption{Comparison of measured low temperature sound absorption data at 2.8 kHz\cite{rau} in a-B$_2$O$_3$ with the soft potential calculation (parameters see Table I).}
\end{figure}
%%%%%%%%%%%%%%%%%%%%% end figure %%%%%%%%%%%%%%%%%%%%%%%%%%%%%%%%%%%%%%%

%%%%%%%%%%%%%%%%%%%%% begin figure %%%%%%%%%%%%%%%%%%%%%%%%%%%%%%%%%%%%%
\begin{figure}[t]
\hspace{-0cm} \vspace{0cm} \epsfig{file=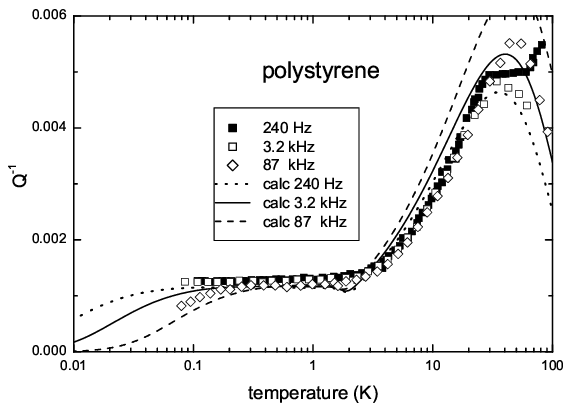,width=7 cm,angle=0} \vspace{0cm} \caption{Comparison of measured low temperature sound absorption data at 240 Hz and 3.2 kHz \cite{nittke}, and at 87 kHz \cite{topp} in polystyrene with the soft potential calculation (parameters see Table I).}
\end{figure}
%%%%%%%%%%%%%%%%%%%%% end figure %%%%%%%%%%%%%%%%%%%%%%%%%%%%%%%%%%%%%%%

%%%%%%%%%%%%%%%%%%%%% begin figure %%%%%%%%%%%%%%%%%%%%%%%%%%%%%%%%%%%%%
\begin{figure}[b]
\hspace{-0cm} \vspace{0cm} \epsfig{file=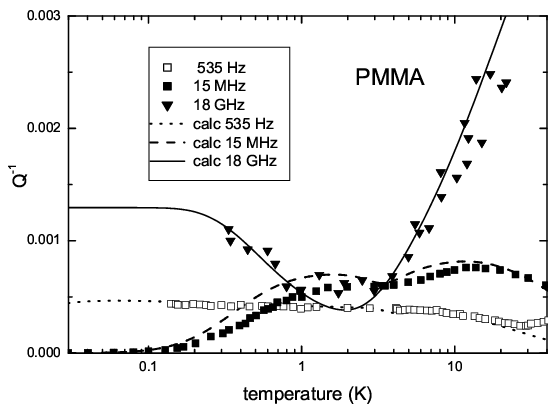,width=7 cm,angle=0} \vspace{0cm} \caption{Comparison of measured low temperature sound absorption data at 535 Hz \cite{nittke}, 15 MHz \cite{federle} and 18 GHz \cite{schmidt} in polymethylmethacrylate with the soft potential calculation, using eq. (\ref{lin}) with a=0.14 (other parameters see Table I).}
\end{figure}
%%%%%%%%%%%%%%%%%%%%% end figure %%%%%%%%%%%%%%%%%%%%%%%%%%%%%%%%%%%%%%%

%%%%%%%%%%%%%%%%%%%%% begin figure %%%%%%%%%%%%%%%%%%%%%%%%%%%%%%%%%%%%%
\begin{figure}[t]
\hspace{-0cm} \vspace{0cm} \epsfig{file=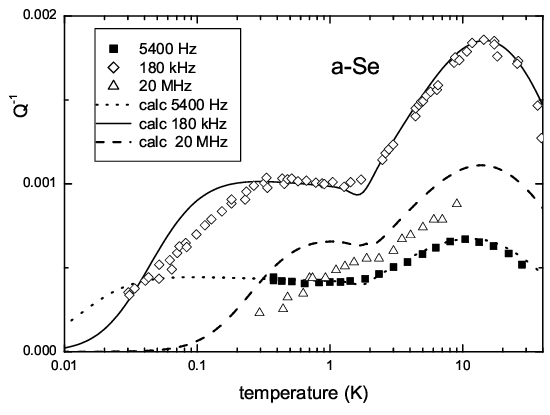,width=7 cm,angle=0} \vspace{0cm} \caption{Comparison of measured low temperature sound absorption data at 5.4 kHz \cite{liu}, 180 kHz \cite{keil}, and at 20 MHz \cite{duq} in amorphous selenium with the soft potential calculation, using eq. (\ref{lin}) with a=0.07 (parameters see Table I, but note that the 180 kHz data are calculated with a factor 2.1 larger  $C$).}
\end{figure}
%%%%%%%%%%%%%%%%%%%%% end figure %%%%%%%%%%%%%%%%%%%%%%%%%%%%%%%%%%%%%%%

%%%%%%%%%%%%%%%%%%%%% begin figure %%%%%%%%%%%%%%%%%%%%%%%%%%%%%%%%%%%%%
\begin{figure}[b]
\hspace{-0cm} \vspace{0cm} \epsfig{file=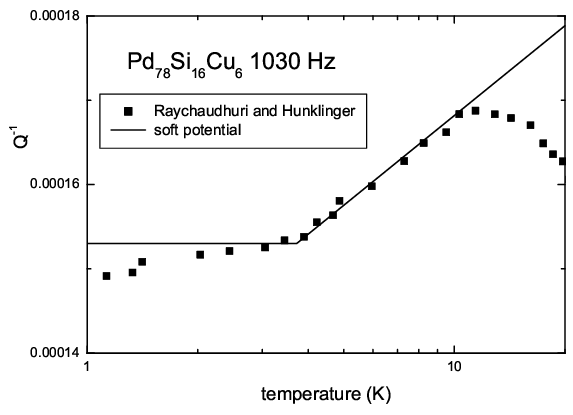,width=7 cm,angle=0} \vspace{0cm} \caption{Low temperature sound absorption data at 1030 Hz \cite{ray} in the metallic glass PdSiCu, showing the crossover from the tunneling plateau to classical relaxation at 3.7 K, implying $W/k_B=3.2$ K.}
\end{figure}
%%%%%%%%%%%%%%%%%%%%% end figure %%%%%%%%%%%%%%%%%%%%%%%%%%%%%%%%%%%%%%%

This interpretation is further supported by the measurements of resonant absorption (see eq. (\ref{restun})) of Golding, Graebner and Schutz \cite{golding} at 500 MHz, which yield again low values of $C$, $C_l=1.9\cdot 10^{-4}$ and $C_t=2.0\cdot 10^{-4}$, compatible within experimental error with the plateau height at lower frequencies in Fig. 4. 

The same discrepancy with the tunneling model predictions, but agreement with the soft potential predictions, is observed for vitreous germania in Fig. 5. The plateau of the 6.3 kHz data \cite{rau} in Fig. 5 (a) agrees with the tunneling prediction, but the plateau of longitudinal ultrasonic data \cite{laermans,pinog} is markedly higher, in agreement with the soft potential calculation. 

Again, the $C_l$-value determined from the 80 MHz sound velocity change in Fig. 5 (b), due to unsaturated resonant tunneling at low temperature is only compatible with the measured plateau height in the same sample \cite{laermans} in the soft potential model, not in the tunneling model. The tunneling model fits yield the same density of states in both measurements, but a markedly stronger coupling constant in the absorption measurement, a result which according to the authors \cite{laermans} has also been found in ultrasonic measurements of the sound absorption and the sound velocity change of neutron-irradiated quartz \cite{laermans1} and in vitreous silica \cite{bartell1}.

These examples show that one can trust the tunneling parameters obtained from sound velocity changes at all frequencies, but not those obtained from the tunneling plateau at higher frequencies. 

In the next two examples, there are only low frequency data, with again good agreement between data and soft potential prediction everywhere, B$_2$O$_3$-data \cite{rau} at 2.8 kHz in Fig. 6 and polystyrene data at 240 Hz and 3.2 kHz \cite{nittke} as well as at 87 kHz \cite{topp} in Fig. 7. The B$_2$O$_3$- and polystyrene data are again well described by the soft potential parameters of the specific heat and conductivity data \cite{ramos}.

In these four examples, one has obviously an essentially constant $P(D_1,D_2)$-region extending beyond the tunneling regime into the relaxation regime. One can recognize these glasses by the validity of the simple approximation \cite{ramos} at low enough frequency, a sudden change of the slope of the damping $Q^{-1}$ {\it versus} temperature, from a constant $Q^{-1}$ in the tunneling region to the $T^{3/4}$-slope of the classical relaxation, which sets in at the crossover temperature $T_c$.     

There are some glasses, however, where the slope at the onset to relaxation is smaller than the predicted one. In the polymer polymethylmethacrylate (PMMA) the onset disappears completely, at least at low frequency (see Fig. 8).

In order to fit the PMMA data in Fig. 8 (535 Hz-data \cite{nittke} 15 MHz-data \cite{federle} and 18 GHz data \cite{schmidt}), it is not enough to assume a quadratic decrease of $P(D_1,D_2)$ with $D_2$; one needs a linear dependence of the form
\begin{equation}\label{lin}
	P(D_1,D_2)=P_s\exp(aD_2).
\end{equation}

In the 535 Hz data \cite{nittke} in Fig. 8, this linear decrease does indeed appear already in the tunneling region; the tunneling plateau is decidedly skew. Of the fitted soft potential parameters \cite{ramos}, only $\Lambda_l$ and $\Lambda_t$ can be used; $C$ has to be changed to a 2.4 times higher value (thus coming close to the one in polystyrene) and $W$ to a 1.55 times higher value (of course, if $P(D_1,D_2)$ is strongly $D_2$-dependent, one cannot expect reliable soft potential values from a fit \cite{ramos} which assumes a constant $P(D_1,D_2)$). 

\begin{table}[htbp]
	\centering
		\begin{tabular}{|c|c|c|c|c|c|c|c|c|}
\hline
subst.                    &$T_g$  &$\rho$   &v$_l$ &v$_t$ &  $W/k_B$   &$P_s/\rho$  &  C      &  b   \\
\hline
                          & K     &kg/m$^3$ & m/s  & m/s  &    K       &10$^{19}$/kg&10$^{-4}$&      \\
\hline
SiO$_2$                   & 1473  &2200     & 5800 & 3800 &  3.9       &  6.3       &    2.6  &  1.8 \\
GeO$_2$                   & 818   &3600     & 3680 & 2410 &  3.8       &  3.9       &    2.2  &  0.06\\
B$_2$O$_3$                & 554   &1810     & 3390 & 1870 &  2.65      &  2.3       &    3.1  &  0.46\\
PS                        & 375   &1050     & 2800 & 1500 &  1.9       &  11.       &    7.1  &  0.40\\
\hline
                          &       &         &      &      &            &            &         &  a   \\
\hline
PMMA                      & 379   &1180     & 3150 & 1570 &  3.5       &  14.       &    8.3  & 0.14 \\
Se                        & 308   &4300     & 2000 & 1050 &  1.6       &  0.84      &    4.4  & 0.07 \\
PdSiCu                    & 550   &10400    & 4790 & 2080 &  3.2       &  0.40      &    1.0  &      \\
\hline                                                                       
		\end{tabular}
	\caption{Sound wave and soft potential parameters for seven glasses (PS=polystyrene, PMMA=polymethylmethacrylate). For the first four glasses, the soft potential parameters are taken from ref. \cite{ramos}, but in the last three, described by eq. (\ref{lin}), $W$ and $C$ are adapted to the sound absorption data.}
	\label{tab:rse1}
\end{table}

At 15 MHz, one recovers the relaxation onset both in the experiment \cite{federle} and in the calculation, but with a lower slope than in the usual case. At the high frequency of 18 GHz, where one samples lower barriers, the relaxation rise is clearly seen. At low temperature, one has to add the resonant tunneling absorption of eq. (\ref{restun}). One needs $D_2=-5$ to reach the tunnel splitting $\Delta_0/k_B=0.85$ K for resonance, so with the weakening factor $a=0.14$ of Table I the effective $C$ is only half of the one in Table I, in good agreement with the value $C=4\cdot10^{-4}$ determined in the original work \cite{schmidt}.

In this case, all three measurements covering the whole accessible frequency range are well described by the soft potential model modification of eq. (\ref{lin}).

In the case of amorphous selenium in Fig. 9, there is also a linear decrease of $P(D_1,D_2)$, but it is a factor of two weaker than in PMMA. Again, one has to change the soft potential parameters \cite{ramos}, the $W$ to a factor 4/3 higher value. The factor for $C$ is different for the two measurements; the measurements at 5400 Hz \cite{liu} and 20 MHz \cite{duq} require a $C$ of 4.4 10$^{-4}$, the one at 180 kHz \cite{keil} a $C$ of 9.3 10$^{-4}$.

There is another measurement \cite{calem} at 228 Hz which shows an even lower $C$ as the 5400 Hz data \cite{liu} and the 20 MHz data \cite{duq}, so there must be a calibration error by a factor of 2.1 in the 180 kHz measurement. The factor is too large to be ascribed to annealing effects, in selenium \cite{liu} of the order of twenty percent.

The last example is the metallic glass PdSiCu at 1030 Hz \cite{ray} in Fig. 10, again with a rather small slope, evaluated in the simple crossover scheme \cite{ramos}. The crossover determines $W/k_B$ at 3.2 K, other data \cite{esqui} at 470 Hz put the value at 2.6 K. The importance of this last metallic glass example lies in the connection to numerical results, because metallic glasses are the real counterpart of the binary glasses, for which most of the numerical results have been obtained so far. 

A very recent numerical study \cite{edan} has determined the soft potential parameters for a binary glass with a repulsive $1/r^{10}$-potential and scaled them via the shear modulus, the atomic mass and the atomic volume to PdSiCu. The obtained value $W/k_B$ of 3.2 K is in excellent agreement with the two experimental values reported here. Also, the scaled value for the tunneling model coupling constant $\gamma_t=0.37$ eV is close to the value 0.4 eV determined in the pioneering resonant tunneling experiment \cite{gold} in PdSiCu.    

\subsection{Sound absorption at higher temperatures}

While the tunneling plateau and the crossover to classical relaxation are universal glass features, each glass develops its own individuality at higher temperatures, showing that different glasses have different energy landscapes at higher barriers. This is well known from many investigations and is seen not only in the mechanical response, but also in other relaxation responses, of which the most prominent is the dielectric one \cite{broad}.

The usual notation is to denote the strong relaxation peak at the glass transition as $\alpha$-peak. One often finds a weaker peak just below the $\alpha$-peak, which merges with the $\alpha$-peak at higher temperatures. This is called the $\beta$-peak. If there are more peaks at lower temperatures, one continues the notation with $\gamma$-peak, $\delta$-peak and so on.

The emphasis of the present work is on the glass phase, with a structure and an energy landscape which can still be considered to be temperature-independent, in terms of the Gilroy-Phillips model \cite{gilroy} a glass with a temperature-independent barrier density $l(V)$. The main aim of this part of the work is to show that in addition to the individual landscape properties one sees again a universal feature, namely the frozen fast $t^\beta$-Kohlrausch tail of the $\alpha$-process (this Kohlrausch $\beta$ between 0.3 and 0.6 has nothing to do with the unfortunate denomination of the $\beta$-process and is in fact best seen in glasses without a $\beta$-peak).

For these higher barriers, one can replace the barrier-dependent $\tau_0$ of eq. (\ref{tau0}) by the usual assumption $\tau_0=10^{-13}$ s.  

%%%%%%%%%%%%%%%%%%%%% begin figure %%%%%%%%%%%%%%%%%%%%%%%%%%%%%%%%%%%%%
\begin{figure}[t]
\hspace{-0cm} \vspace{0cm} \epsfig{file=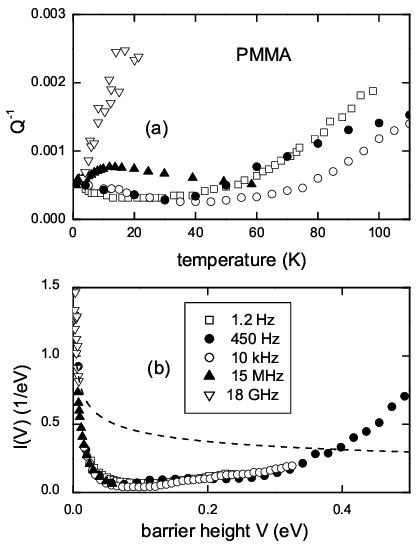,width=7 cm,angle=0} \vspace{0cm} \caption{(a) Measured sound absorption data in polymethylmethacrylate at 1.2 Hz \cite{sinnott}, 450 Hz \cite{geis}, 10 kHz \cite{crissman}, 15 MHz \cite{federle} and 18 GHz \cite{schmidt} (b) barrier density $l(V)$ calculated from the data in (a) with eq. (\ref{lvq}) and $\tau_0=10^{-13}$ s; the dashed line indicates the soft potential expectation.}
\end{figure}
%%%%%%%%%%%%%%%%%%%%% end figure %%%%%%%%%%%%%%%%%%%%%%%%%%%%%%%%%%%%%%%

%%%%%%%%%%%%%%%%%%%%% begin figure %%%%%%%%%%%%%%%%%%%%%%%%%%%%%%%%%%%%%
\begin{figure}[b]
\hspace{-0cm} \vspace{0cm} \epsfig{file=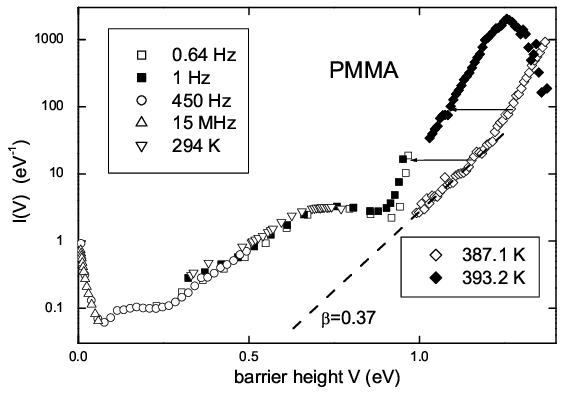,width=7 cm,angle=0} \vspace{0cm} \caption{Barrier density $l(V)$ in polymethylmethacrylate calculated from torsion pendulum data at 0.64 Hz \cite{schwarzl} and at 1 Hz \cite{heijboer}, from sound absorption data at many different frequencies at 294 K \cite{read}, and from $J_r(t)$-data \cite{plazek-pmm} in the undercooled liquid at 387.1 K and 393.2 K.}
\end{figure}
%%%%%%%%%%%%%%%%%%%%% end figure %%%%%%%%%%%%%%%%%%%%%%%%%%%%%%%%%%%%%%%

%%%%%%%%%%%%%%%%%%%%% begin figure %%%%%%%%%%%%%%%%%%%%%%%%%%%%%%%%%%%%%
\begin{figure}[t]
\hspace{-0cm} \vspace{0cm} \epsfig{file=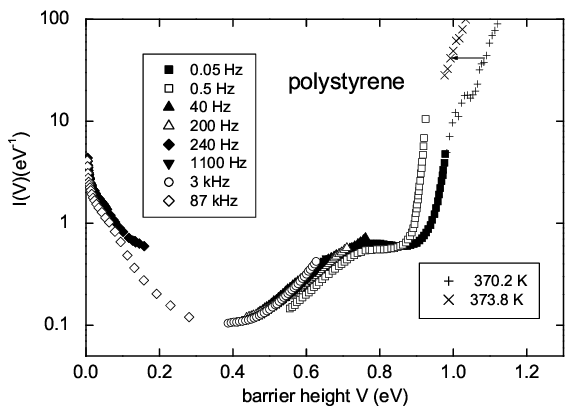,width=7 cm,angle=0} \vspace{0cm} \caption{Barrier density $l(V)$ in polystyrene calculated from the low temperature data at 240 Hz \cite{nittke} and 87 kHz \cite{topp}, a collection of mechanical data at 0.05, 0.5, 40, 200, 1100 and 3000 Hz up to the glass temperature in the same sample \cite{illjenck}, and from $J_r(t)$-data \cite{plazek-ps} in the undercooled liquid at 370.2 K and 373.8 K.}
\end{figure}
%%%%%%%%%%%%%%%%%%%%% end figure %%%%%%%%%%%%%%%%%%%%%%%%%%%%%%%%%%%%%%%

Fig. 11 shows the evaluation of shear data in PMMA below 100 K (1.2 Hz-data \cite{sinnott}, 450 Hz-data \cite{geis}, 10 kHz-data \cite{crissman}, 15 MHz-data \cite{federle} and 18 GHz Brillouin data \cite{schmidt}, the latter two the same data as in Fig. 8) . Fig. 11 (a) shows the data, Fig. 11 (b) the calculated barrier density, together with the soft potential expectation. $l(V)$ falls first below the soft potential expectation, then rises again as one begins to approach the strong $\beta$-peak of PMMA. It is amazing to see how well the Gilroy-Phillips model brings the divergent data of Fig. 11 (a) into a consistent energy landscape picture, with such a crude approximation as eq. (\ref{lvq}). The agreement corroborates the Gilroy-Phillips assumption of a constant distribution of asymmetries $n(V,\Delta)$ around the value $\Delta=0$. 

Fig. 12 extends the PMMA picture up to the glass temperature, using temperature-dependent torsion pendulum data at 0.64 Hz \cite{schwarzl} and at 1 Hz \cite{heijboer}. These data start around 70 K at a barrier of about 0.2 eV, where $GJ'(\omega)$ has risen to the value 1.03, so one can integrate further starting from this value.

The torsion pendulum data show the strong secondary relaxation peak of PMMA at 0.7 eV in good agreement with each other and with a collection of data \cite{read} between 0.01 Hz and 10 MHz at room temperature, 294 K. At the upper end of the peak, $GJ'(\omega)$ has grown to 1.8.

But PMMA is a polymer, where the true viscous flow sets in many decades later in time, depending on the chain length. The rise after the secondary relaxation peak consists entirely of reversible relaxations. What does happen, though, is the glass transition, the transition from a frozen glass to an undercooled liquid in thermal equilibrium.

The polymer community is used to this phenomenon. They call the glass transition "segmental relaxation" to indicate that it happens on short pieces of the polymer, while the polymer as a whole equilibrates much later. But this separation of thermal equilibration and viscous flow does not exist in a normal glass.  

The barrier distribution $l(V)$, frozen in the glass, begins to increase rapidly with increasing temperature in the undercooled liquid. This is shown in Fig. 12 by the comparison of the distributions derived with eq. (\ref{jtlv}) from $J_r(t)$-data \cite{plazek-pmm} at 387.1 and 393.2 K, using $G=2.09$ GPa at 387.1 K and $G=2.03$ GPa at 393.2 K, values determined from transverse Brillouin data \cite{kruger} of PMMA.

The evaluation of the torsion pendulum data in Fig. 12 ends at 393 K, shortly above the glass transition. It is seen that, at this temperature, they have already adapted to the higher $l(V)$ of the liquid and do no longer reflect the barrier density of the glass.

What does reflect the barrier density of the glass, however, is the one for the equilibrium liquid at 387.1 K (or maybe 1 or 2 K higher, depending on the cooling rate, with which the glass was frozen). This shows an exponential rise with $\exp(0.37V/k_BT)$, leading to a Kohlrausch $t^\beta$ time dependence of the shear relaxation with a Kohlrausch $\beta=0.37$.

In the undercooled liquid, the Kohlrausch tail at a given barrier height increases by a factor of seven between 387.1 and 393.2 K (more precisely, the barrier of a given Kohlrausch process decreases according to the shift factor between the two temperatures \cite{plazek-pmm}, see the two arrows in Fig. 12). Since some of the measurements of the secondary relaxation peak with different frequencies \cite{read} were also done close to the glass transition (at 373.2 K) and showed practically no change of its height, one can conclude that the drastic rise remains limited to the Kohlrausch tail.

Fig. 13 shows the same picture in another well-studied polymer, polystyrene, combining the low-temperature data \cite{nittke,topp} with a collection of mechanical data at six frequencies between 0.05 Hz and 3 kHz at higher temperatures \cite{illjenck}, and with the $J_r(t)$-measurements \cite{plazek-ps} in the equilibrium liquid, evaluated with $G=1.18$ GPa from Brillouin data \cite{strube}.

In this case, the resulting barrier distribution between 0 and 0.1 eV is supported by light scattering data \cite{soko1} evaluated in terms of the Gilroy-Phillips model and related to the mechanical data \cite{topp} many years ago. 

%%%%%%%%%%%%%%%%%%%%% begin figure %%%%%%%%%%%%%%%%%%%%%%%%%%%%%%%%%%%%%
\begin{figure}[t]
\hspace{-0cm} \vspace{0cm} \epsfig{file=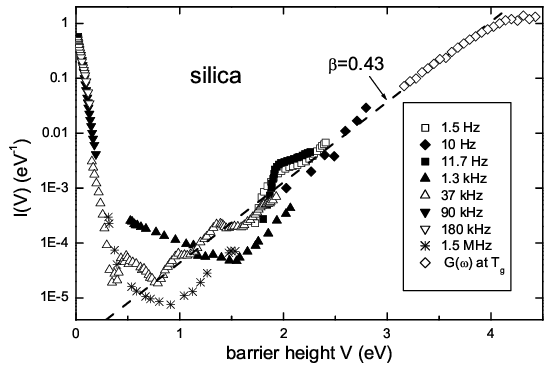,width=7 cm,angle=0} \vspace{0cm} \caption{Barrier density $l(V)$ in vitreous silica from measurements at 1.5 and 11.7 Hz \cite{kirby}, 10 Hz \cite{brueckner}, 1.3 kHz \cite{deeg}, 37 kHz \cite{marx}, 90 kHz \cite{topp}, 180 kHz \cite{keil}, 1.5 MHz \cite{fraser}, and from $G(\omega)$-data \cite{mills} in the undercooled liquid at 1449 K.}
\end{figure}
%%%%%%%%%%%%%%%%%%%%% end figure %%%%%%%%%%%%%%%%%%%%%%%%%%%%%%%%%%%%%%%

%%%%%%%%%%%%%%%%%%%%% begin figure %%%%%%%%%%%%%%%%%%%%%%%%%%%%%%%%%%%%%
\begin{figure}[b]
\hspace{-0cm} \vspace{0cm} \epsfig{file=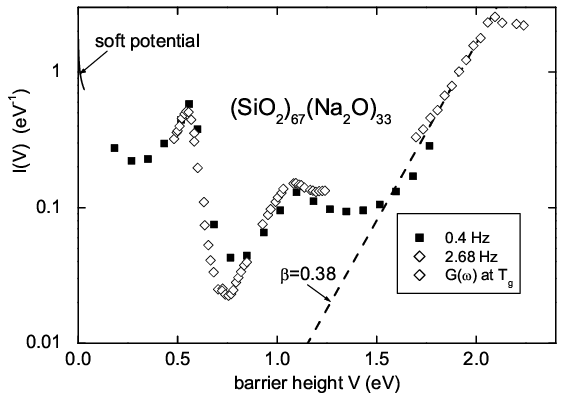,width=7 cm,angle=0} \vspace{0cm} \caption{Barrier density $l(V)$ in (SiO$_2$)$_{67}$(Na$_2$O)$_{33}$ from measurements at 0.4 Hz \cite{rindone}, 2.68 Hz \cite{forry}, and from $G(\omega)$-data \cite{mills} in the undercooled liquid at 728 K. The short line at small barriers is the soft potential expectation from the parameters determined for a sodium silicate glass with a slightly lower sodium content \cite{ramos}.}
\end{figure}
%%%%%%%%%%%%%%%%%%%%% end figure %%%%%%%%%%%%%%%%%%%%%%%%%%%%%%%%%%%%%%%

%%%%%%%%%%%%%%%%%%%%% begin figure %%%%%%%%%%%%%%%%%%%%%%%%%%%%%%%%%%%%%
\begin{figure}[t]
\hspace{-0cm} \vspace{0cm} \epsfig{file=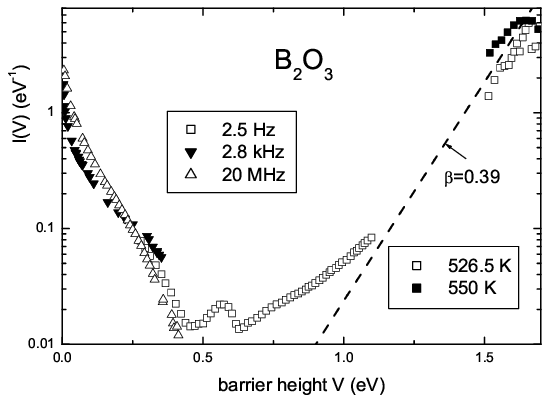,width=7 cm,angle=0} \vspace{0cm} \caption{Barrier density $l(V)$ in B$_2$O$_3$ from measurements at 2.5 Hz \cite{karsch}, 2.8 kHz \cite{rau}, 20 MHz \cite{kurkjian} and from $J_r(t)$-data \cite{plazek-bo} in the undercooled liquid at 526.5 and 550 K.}
\end{figure}
%%%%%%%%%%%%%%%%%%%%% end figure %%%%%%%%%%%%%%%%%%%%%%%%%%%%%%%%%%%%%%%

%%%%%%%%%%%%%%%%%%%%% begin figure %%%%%%%%%%%%%%%%%%%%%%%%%%%%%%%%%%%%%
\begin{figure}[b]
\hspace{-0cm} \vspace{0cm} \epsfig{file=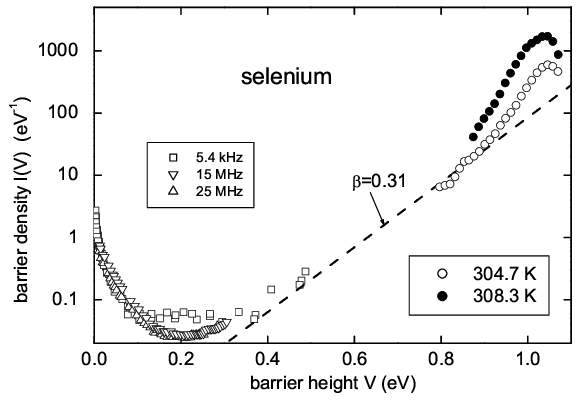,width=7 cm,angle=0} \vspace{0cm} \caption{Barrier density $l(V)$ in selenium from measurements at 5.4 kHz \cite{liu}, 15 and 25 MHz \cite{pino} and from $J_r(t)$-data \cite{roland} in the undercooled liquid at 304.7 and 308.3 K.}
\end{figure}
%%%%%%%%%%%%%%%%%%%%% end figure %%%%%%%%%%%%%%%%%%%%%%%%%%%%%%%%%%%%%%%

As it turns out, the Kohlrausch tail appears also in non-polymeric glasses, only that there it is directly followed by the viscous flow which ends all reversible processes. This is shown for vitreous silica in Fig. 14, displaying many data from the literature evaluated in a previous publication \cite{philmag2002} in terms of $f(V)$. The evaluation in terms of $l(V)$ in Fig. 14 is not very different, because significant deviations of $GJ'$ from one begin only to appear close to the glass transition. In fact, the $l(V)$ of silica within experimental error consists of nothing else than low temperature relaxations and Kohlrausch tail, with a Kohlrausch $\beta$ of 0.43, already visible in many of the glass measurements \cite{keil,topp,kirby,brueckner,deeg,marx,fraser} and continued in the equilibrium $G(\omega)$ at 1449 K \cite{mills}. For the evaluation of the $G(\omega)$-data, one takes the viscosity $\eta=3014$ GPas from a fit \cite{asyth1} of the data \cite{mills} and uses $G=35$ GPa from a transverse Brillouin scattering measurement \cite{dardy}. The integral over all reversible relaxations, the total recoverable compliance $J_0$, yields $GJ_0=2.2$.

Again, the low-barrier part of the barrier distribution in Fig. 14 was determined much earlier from light scattering data \cite{soko2}, and shown to be the same as the one from mechanical data for barrier heights between zero and 0.1 eV. 

If one puts alkali metals into silica, secondary relaxations appear and the glass temperature is drastically lowered. This is shown in Fig. 15 for (SiO$_2$)$_{0.67}$(Na$_2$O)$_{0.33}$. The torsion pendulum measurements at 0.4 Hz \cite{rindone} and at 2.6 Hz \cite{forry} show two secondary relaxation peaks, which together bring $GJ'$ up to 1.3. It follows that the real $G$ at the glass temperature 728 K of the equilibrium $G(\omega)$ \cite{mills} is a factor of 1.3 larger than the value $G=14.5$ GPa fitted \cite{asyth1} to the equilibrium data. Therefore the calculation of the equilibrium $l(V)$-data at 728 K in Fig. 15 was done with $G=18.9$ GPa and $\eta=1415$ GPas, the fitted viscosity \cite{asyth1}. The data show a Kohlrausch tail with $\beta=0.38$.

The next case, B$_2$O$_3$ in Fig. 16, has again a relatively small $l(V)$ in the intermediate region between low barrier relaxations and Kohlrausch tail. In the glass phase, the $l(V)$ is obtained from torsion pendulum data at 2.5 Hz \cite{karsch}, the 2.8 kHz data \cite{rau} already shown in Fig. 6, and 20 MHz data \cite{kurkjian}. The equilibrium $l(V)$-data were determined from $J_r(t)$-measurements \cite{plazek-bo} at 526.5 and 550 K, with $G=6.5$ and 6.3 GPa, respectively, from a transverse Brillouin measurement \cite{grimsditch}. 

Note that the factor between the $l(V)$ at the two equilibrium temperatures is much smaller than in the PMMA case of Fig. 12, though the temperature difference is much larger. This reflects the much smaller fragility of B$_2$O$_3$ (the fragility index $m$ is 32 in B$_2$O$_3$ and 145 in PMMA \cite{bohmer}); the temperature dependence of the Kohlrausch tail in the liquid is much smaller in liquid B$_2$O$_3$ than in liquid PMMA.

In Fig. 16, the 20 MHz-data \cite{kurkjian} extend from low barriers up to 0.4 eV. In this barrier region, one can reduce $l(V)$ dramatically with a small OH content, as shown in the same paper. But as $l(V)$ is small to start with, the effect is not included here.

The last example is selenium in Fig. 17, with $l(V)$ from paddle oscillator data \cite{liu} at 5.4 kHz and longitudinal ultrasonic data \cite{pino} at 15 and 25 MHz in the glass phase, another case where uniaxial and shear distortion provide the same $l(V)$. The equilibrium liquid values are from $J_r(t)$-data \cite{roland} at 304.7 and 308.3 K, evaluated with $G=3.4$ and 3.3 GPa, respectively, values taken from an ultrasonic determination \cite{kozhev} which shows excellent agreement with an earlier one \cite{galli}.

This is again a case where one finds essentially only low-barrier relaxations and Kohlrausch tail, in this substance $\beta=0.31$, close to the Andrade value of 1/3. Since selenium is a short polymer, one again finds a crossover to a polymer-like behavior at long relaxation times.

\section{Discussion and conclusions}

\subsection{Main results}

This is the first systematic investigation of the energy landscape in glasses from low to high barriers on the basis of mechanical relaxations, connecting the viscous flow setting in at the glass transition with the tunneling states at low temperature. One finds a satisfactory description of the sound absorption in glasses, at low temperatures with the soft potential model \cite{klinger,bggprs,parshin,ramos,ramos1,ramos2,schober} and at higher temperatures with the Gilroy-Phillips model \cite{gilroy,burel} in terms of a temperature-independent barrier density which freezes in at the glass transition.

The paper demonstrates the importance of linking sound absorption in real glasses to the increasing flood of new numerical work on frozen glasses \cite{lebo} and its connection to the glass transition \cite{bertun,berthier}, as detailed in the following section IV. B.

The combination of two phenomenological models to describe the sound absorption in glasses allows for a quantitative check of the energy landscape idea, in particular with respect to the question whether there is a constant distribution of asymmetries around the asymmetry zero in the energy landscape. Surprisingly, the answer from the sound absorption data evaluated here is yes for all temperatures, while the earlier investigations with the much more powerful dielectric technique \cite{olsen,gainaru,gainaru2} limit their yes to temperatures below 0.6 $T_g$, because of the strong temperature dependence of the barrier density at the Kohlrausch tail. But there is no such strong temperature dependence of the Kohlrausch tail in the mechanical data evaluated here. The present results indicate that the strong temperature dependence \cite{olsen,gainaru,gainaru2} in the glass phase close to $T_g$ is a peculiarity of hydrogen-bonded glass formers \cite{oh1,oh2}. The issue is discussed in more detail in Section IV. C.

In all six investigated glass formers the barrier density at the barrier height zero is relatively high, comparable to the one of the reversible Kohlrausch tail at the point where it crosses over into the irreversible flow at the glass transition. This result is obviously related to the universality of tunneling states in glasses \cite{phillips,hunk,kauz}. 

From the tunneling states, the barrier density decreases rapidly with increasing barrier height. In the two examples without secondary relaxation peak, vitreous silica and selenium, there is only this decrease and the subsequent increase of the Kohlrausch tail. In these two cases, the Kohlrausch tail does already appear when $GJ'$ is only a few percent away from its starting value 1. This excludes any explanation of the strong Kohlrausch rise in terms of the interaction between different relaxing domains, a concept proposed in several papers \cite{bu2004,bu2009,bu2011} by one of us. It rather supports a Shear Transformation Zone treatment of the highly viscous flow \cite{falk,john,bu2022} which identifies the Kohlrausch processes with cooperatively rearranging Eshelby regions \cite{eshelby} resulting from a combination of several soft modes, together leading to a new sheared stable structure. In this picture, the high density of soft vibrations, tunneling and low barrier relaxational modes is necessary to enable the highly viscous flow, consistent with the findings reported here.

A gratifying quantitative result is the resolution of the discrepancy \cite{topp} between the height of the tunneling plateau at low and high frequencies. The soft potential treatment of the present paper shows that the higher tunneling plateau at higher frequencies is due to the approaching crossover to vibrations.

\subsection{Connections to numerical work}

The last five years brought important new numerical developments \cite{lebo,berthier}. The first was a dedicated study of the localized vibrational soft modes in simple glasses \cite{le1,le2,manning,le3,corein,mizuno,wang,le4,mizuno2,proca,edan} with the vibrational density of states $g(\omega)\propto\omega^4$ ($\omega$ frequency) predicted by the soft potential model \cite{bggprs,parshin,ramos,schober} and exhibiting the strong positive fourth order term $D_4$ in the mode potential \cite{le1,le3,le4} which the soft potential model needs for a common description of tunneling states and vibrations. The second was the swap mechanism for simple liquids \cite{swap}, which enables the numerical cooling of simple liquids down to temperatures which are even lower than the glass temperature of real liquids. The application of the swap mechanism to undercooled liquids revealed the central role of the soft localized modes for the understanding of the mode coupling transition \cite{coslovich} and documented a strong decrease of the number of these soft modes in the glass phase with decreasing glass temperature \cite{wang,bertun}.

Three of these new papers \cite{corein,mizuno2,edan} corroborated an earlier numerical result \cite{corei}, namely the finding of an unstable core of the soft vibrational modes. The small positive force constant results from the compensation of the negative restoring force of the core by the positive restoring force of the stable surroundings.

There is a very recent numerical result \cite{le4}, showing that one has double well potentials with a barrier density proportional to $V^{1/4}$, where $V$ is the barrier height. The work is based on an earlier ingenious characterization \cite{le2,le5} of the soft modes in terms of eigenvectors defined over the fourth and third terms of the mode potential. This characterization led to the surprising result that the eigenvectors defined over the fourth order term are very close to the usual second order ones, allowing one to get rid of the influence of the hybridization between localized modes and phonons.

In the soft potential model, the double well potentials are due to modes with a negative restoring force $D_2$, as long as the absolute value of the linear potential coefficient $D_1$ stays below the limit of eq. (\ref{d1lim}). For a constant density of modes in the $D_1,D_2$-plane, this leads to the barrier density increasing with $V^{1/4}$, in agreement with the numerical result \cite{le4}.

Recently the soft potential parameters have been determined for a binary numerical glass \cite{edan}, making use of new numerical results \cite{le6}. Scaling the parameters with the shear modulus, the atomic mass and the atomic volume to metallic glasses, one finds quantitative agreement with the $W$ of PdSiCu determined from the data \cite{ray} in Fig. 10 of the present paper, and with the coupling constant between tunneling states and transverse sound waves in the same material \cite{gold}. It would be worthwhile to extend the comparison to new numerical treatments of network and molecular glasses \cite{le7}.

The concept \cite{bu2022} of cooperatively rearranging regions consisting of several unstable soft mode cores is supported by a very recent numerical result \cite{roying}, showing that below the mode coupling temperature the large single particle displacements occur predominantly within the cooperatively rearranging regions. 

\subsection{Comparison to light scattering and dielectric results}

Our Gilroy-Phillips interpretation of the mechanical data at low barriers is the same as the earlier evaluation of light scattering data in Ca$_{0.4}$K$_{0.6}$(NO$_3$)$_{1.4}$, polystyrene, and polycarbonate \cite{soko1} as well as in vitreous silica \cite{soko2}, as mentioned in connection with our polystyrene and silica results in Section III. B.

Broad band dielectric spectroscopy \cite{broad} is the easiest and most accurate method to study relaxations. It has been extensively applied to relaxations in glasses by the Bayreuth group. The results are summarized in Reference \cite{gainaru} and demonstrated several of the results of the present paper eleven years ago, among them the rise to stronger relaxation at very low barriers and the validity of the Gilroy-Phillips model, demonstrated with an accuracy of which mechanical investigations can only dream.

But there is one central dielectric result \cite{gainaru,gainaru2,olsen} for which one finds no trace in the present survey of mechanical data, namely the strongly temperature-dependent barrier density at the frozen Kohlrausch tail which fails to follow the Gilroy-Phillips predictions. In the dielectric data, it is not always observed. It is clearly seen in glycerol, but it is absent in toluene \cite{gainaru}, a molecule with no hydrogen bonds.

The simplest possible explanation of the temperature dependence $\exp(5T/T_g)$ of the excess wing measured in the glass phase \cite{gainaru} is an average asymmetry of 5$k_BT_g$ of the double-well potentials. This value is close to the average asymmetry of 3.8 $k_BT_g$ needed to explain the intensity rise of the strong secondary relaxation peak in tripropylene glycol after the initial temperature jump of an aging experiment \cite{olsen}.

In the vitreous silica data in Fig. 14 and the selenium data of Fig. 17, there is no such strong temperature dependence. Also, the high quality PMMA data \cite{read} of the secondary relaxation peak show no deviation at all from the Gilroy-Phillips model.  

An asymmetry of 4$k_BT_g$ can be excluded with absolute certainty for the mechanical data. To see this, consider the barrier 0.3 eV in PMMA, measured at 294 K with 10 MHz \cite{read} and at 127 K with 1.2 Hz \cite{sinnott}. This follows from eq. (\ref{taurel}) with $\tau_0=10^{-13}$ s. With $T_g=387$ K, the weakening factor $(\cosh(\Delta/2k_BT))^{-2}$ by the asymmetry 4$k_BT_g$ would be 0.02 for the high frequency and a factor of thousand smaller for the low frequency. The fact that the two frequencies see the same barrier density at the two different temperatures shows that the average asymmetry is zero.

The same conclusion can be drawn from the silica data in Fig. 14. One sees the same barrier density at 1.5 Hz \cite{kirby} and 37 kHz \cite{marx}, though the measurement temperatures of the 1.5 Hz data are only two thirds of the ones for 37 kHz, excluding an average asymmetry larger than a small fraction of $k_BT_g$.   

From NMR measurements \cite{nmr}, it is known that an irreversible relaxation in the primary relaxation peak of glycerol is a motion of many molecules, of which only about 2 percent make large angle (30 to 50 degrees) jumps; the rest make small angle jumps of a few degrees. With this information in mind, it seems possible that the excess wing is due to reversible reorientational jumps of single hydrogen bonds \cite{hb} in strongly asymmetric double-well potentials, similar to the fast reversible hydrogen bond jumps in water \cite{bagchi}, the large dipole moment change providing a signal strength able to compensate the weakening factor from the strong asymmetry.

The imaginary peak of depolarized dynamical light scattering data in glycerol and other hydrogen bonded glass formers lies a factor of three higher than the one of the dielectric data \cite{pabst}. Looking for an explanation of the upwards peak shift from dielectrics to depolarized dynamical light scattering in glycerol, one remembers that for the simple case of isotropic rotational diffusion of a molecular dipole Debye \cite{debye} predicts the peak in $\epsilon''$ at $\omega=2D_r$ ($D_r$ rotational diffusion constant) and Berne and Pecora \cite{pecora} predict the imaginary peak in the depolarized dynamical light scattering at $\omega=6D_r$, a factor of three higher. This explanation of the new depolarized dynamical light scattering data \cite{pabst} is consistent with the one of the NMR data \cite{nmr}.

\subsection{Conclusions}

To summarize, a survey of many sound absorption data in glasses from the literature shows a high density of low barrier relaxations, connected with the universal low temperature glass anomalies, well described by the soft potential model, an extension of the tunneling model to include low barrier relaxations and soft vibrations. The soft potential model explains the rise of the tunneling plateau with increasing frequency which cannot be understood in the tunneling model. It finds strong support from recent numerical work on frozen binary glasses. 

Starting from low barriers, the barrier density decreases toward higher barriers, followed by the region of intermediate barriers which looks different for different glasses, for many glasses containing one or more secondary relaxation peaks. This region and the universal Kohlrausch tail of the highly viscous flow toward higher barriers are found to be describable in terms of a temperature-independent barrier density. According to the Gilroy-Phillips model, this means one has a constant distribution of asymmetries around the value zero in the corresponding asymmetric double-well potentials. This conclusion agrees only partly with earlier dielectric investigations, because it does hold neither for the excess wing of hydrogen bonded glass formers nor for the secondary relaxation peak in tripropylene glycol. The present results suggest that these deviations are due to the breaking of hydrogen bonds in strongly asymmetric double well potentials and are not a generic feature of undercooled liquids. 

The Kohlrausch tail is temperature-independent in the glass, but becomes strongly temperature-dependent in the undercooled liquid, the more so the higher the fragility is. At the glass temperature, its barrier density at the barrier corresponding to the Maxwell time is comparable to the one of the low temperature anomalies, suggesting a connection between the two.

\section*{Acknowledgements}

M.A.R. acknowledges financial support from the Spanish Ministry of Science and Innovation through the ``Mar\'{\i}a de Maeztu''  Programme for Units of Excellence in R\&D (CEX2018-000805-M), as well as from the Autonomous Community of Madrid through program S2018/NMT-4321 (NANOMAGCOST-CM).

\end{document}